\documentclass[11pt,twoside]{article}
\usepackage{asp2004}
\usepackage{psfig}
\usepackage{epsf}
\usepackage{graphics}
\usepackage{lscape}
\def\edcomment#1{\iffalse\marginpar{\raggedright\sl#1\/}\else\relax\fi}
\reversemarginpar

\begin{document}
\title{Magnetically Preferred Solar Longitudes: Reality?}
\author{C. J. Henney}
\affil{National Solar Observatory, Tucson, Arizona, USA}
\author{B. R. Durney}
\affil{2377 Route de Carc\'es, FR-83510 Lorgues, France.}

\begin{abstract}
The observed persistence of specific periodicities detected in time series 
associated with solar surface magnetic activity over several solar 
cycles has led to numerous papers supporting the existence of preferred 
longitudes. Recent analysis of the past 120 years of sunspot number 
data showed that no observed periodicity remained coherent for durations 
greater than two 11-year solar cycles. Here we address the question: Could 
the observed periodicities of solar magnetic signals on time scales of
two decades be the result of a purely stochastic process? We begin to answer 
this by comparing phase coherence between observed periodic signals and 
signals from a model using longitudinally random eruptions. A surprisingly 
non-negligible likelihood is found, approximately 1 in 3, that observed 
periodicities from integrated full-disk solar parameters are a chance 
occurrence for time series on the order of 20 years in duration.
\end{abstract}

\section{Introduction}
The observed solar surface is spatially and temporally variable
due to the dynamic nature of magnetic active regions along with 
differential rotation and meridional flows. Even with such 
dynamics, active regions are not observed to be distributed
randomly but rather magnetic regions tend to form ``complexes
of activity'' \citep[e.g.][]{Bum65,bz90}. Using Kitt Peak Vacuum 
Telescope (KPVT) magnetic synoptic maps,
persistent bands of active ``nests'' were reported by \citet{detoma00} 
to support the idea of a stable pattern in the tachocline for 
sources of emerging flux. The lifetimes of these activity bands were
found to be as long seven solar rotations.

For longer time scales, the temporal coherence of magnetic 
regions is expected to decrease for periods greater than 
the lifetime of the active regions. However, there have been numerous
reports of persistent signals from spectral analysis with the sunspot 
number time series \citep[e.g.][]{Sval75,Bogart82}. Using autocorrelation
analysis with 128 years of sunspot number time series data, \citet{Bogart82}
found a persistent 27.5-day period signal. Additionally, work by \citet{Neu00}
reported a very persistent signal with a well-defined period
of 27.03 $\pm$ 0.002 days, using over 30 years of solar magnetic field
and solar wind measurements. \citet{hen02} 
also detected this signal using photospheric magnetograms (see Figure~1). 
In addition, using phase coherence analysis on the past 120 years of sunspot 
number data, \citet{hen02} showed that no observed periodicity remained coherent 
for durations greater than two solar cycles. Here we estimate the likelihood 
that coherent signals for durations on the order of two solar cycles 
are a result of chance occurrence. 
The model developed and used for this study can operate in several modes
ranging from purely random longitudinal location for the infusion of new flux to
including a specific longitude and rotation rate for new flux at any selected 
latitude. Since the integrated full-disk signal is of interest here, the solar model 
was used in its random longitude mode and does not include detailed solar surface
flux transport \citep[e.g.][]{Schrij03}. The following section outlines the 
magnetic activity model utilized for this preliminary study.

\begin{figure}
\plotone{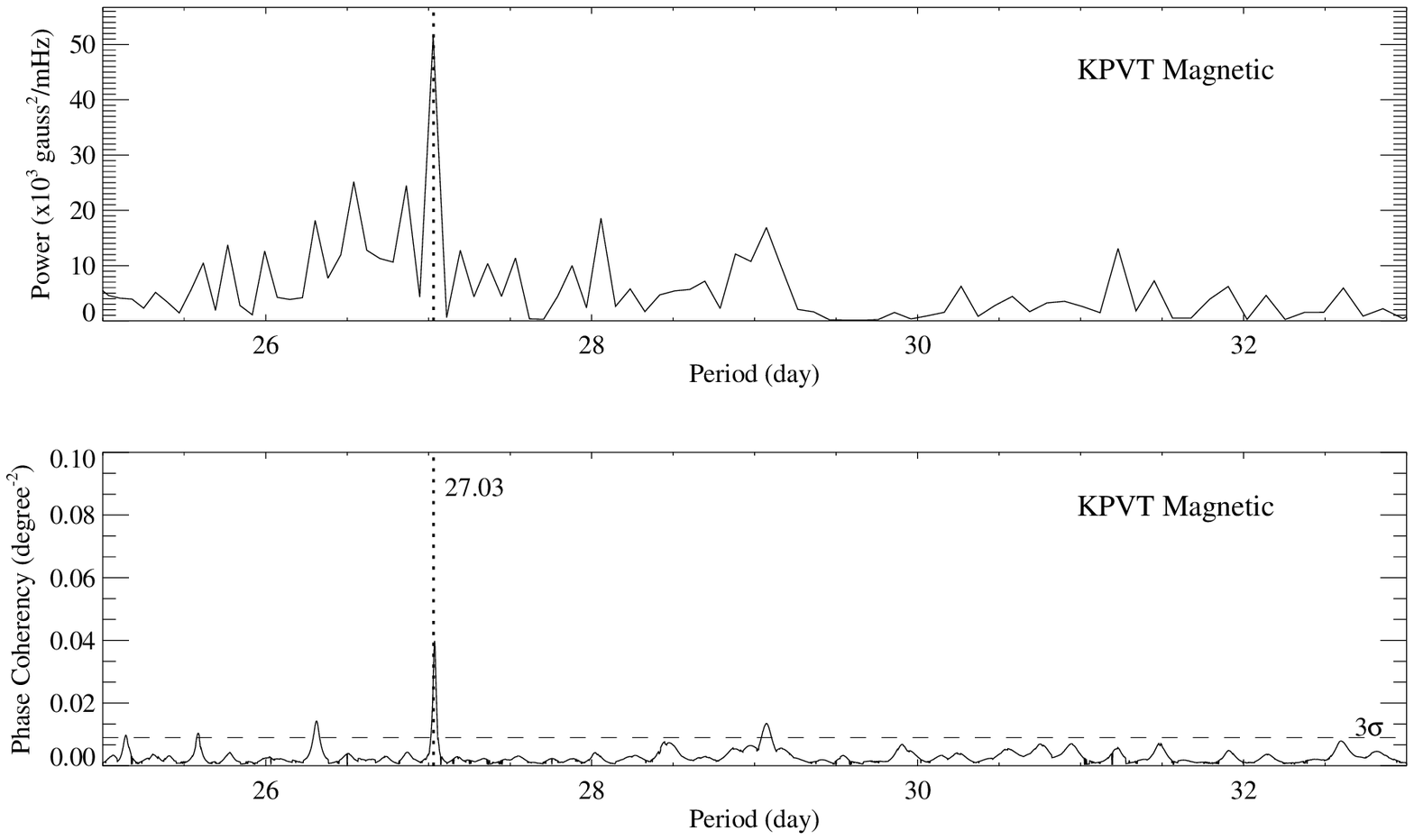}
\plotone{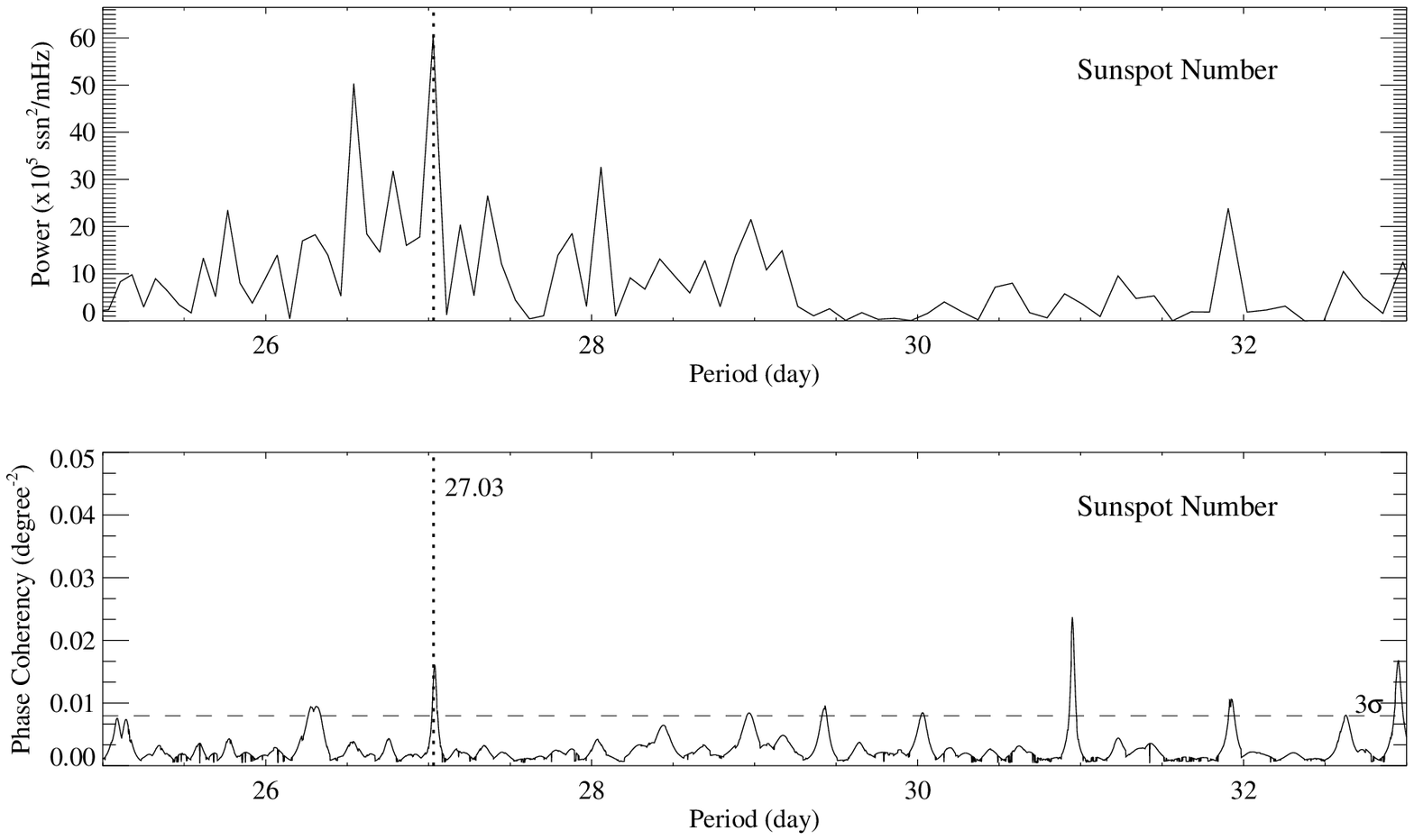}
\caption{Comparison of power and phase coherence spectra 
between Kitt Peak Vacuum Telescope (KPVT) unsigned magnetic flux 
integrated full-disk time series (top two), 
and the international sunspot number (bottom two) time 
series \citep[from][]{hen02}. 
Each time series is sampled at a cadence of 1 day and for the same 8839-day 
interval (1977.2 to 2001.4, in units of fractional years). 
The 3-$\sigma$ noise level for the phase coherence spectra
is delineated by the horizontal dashed lines. The vertical dotted 
line illustrates the 27.03-day period position.}
\end{figure}

\section{Magnetic Activity Model}
The modeling of large-scale solar magnetic activity is achieved in several 
steps. The initial step is to estimate the values of the mean eruption 
strength and decay-time for observed flux within a given latitude band.
These values were determined using the same unsigned magnetic synoptic 
maps created from KPVT full-disk photospheric
measurements that were used by \citet{hen02}. The KPVT synoptic map data 
is described by \citet{Harvey80}. 
The long-term trends of solar magnetic activity are determined for 18 latitude 
bins of the KPVT synoptic magnetic maps. The maps were spatially 
sampled in sine-latitude bins and temporally sampled to create latitudinal 
time series. The temporal gaps are filled with a high-order autoregressive model, 
and each time series is temporally filtered using a non-recursive digital filter 
with a low-band pass between 10 and 60 days. Each time series is then median 
filtered with a window size of 83-days, then finally smoothed by convolving 
a Bartlet function with a window size of 27-days. 

The filtered-flux time series is sampled at a cadence of 10 days, then is 
interpolated to a final cadence of 1 day. Allowing for daily eruptions, the 
model eruption strength and decay-time for each latitude band are 
then estimated by minimizing the standard deviation between the filtered- and 
modeled-flux time series (see {Figure~2}). The eruption strengths shown in Figure~2
vary from approximately $1 \times 10^{22}$ to $3.5 \times 10^{22}$~Mx. 
These values are in agreement
with previous estimates of $1 \times 10^{22}$~Mx and  $2 \times 10^{22}$~Mx 
for observed ephemeral \citep[e.g.][]{HM1979} and active \citep[e.g.][]{NS1966} 
region strengths respectively. The estimated decay periods deviate around 
100 days, which correspond to approximately $1 \times 10^{20}$ to 
$3 \times 10^{20}$~Mx/day. These decay rates are consistent with previous 
estimates of $2 \times 10^{20}$~Mx/day reported by \citet{KH93}.

\begin{figure}[!t]
\plotone{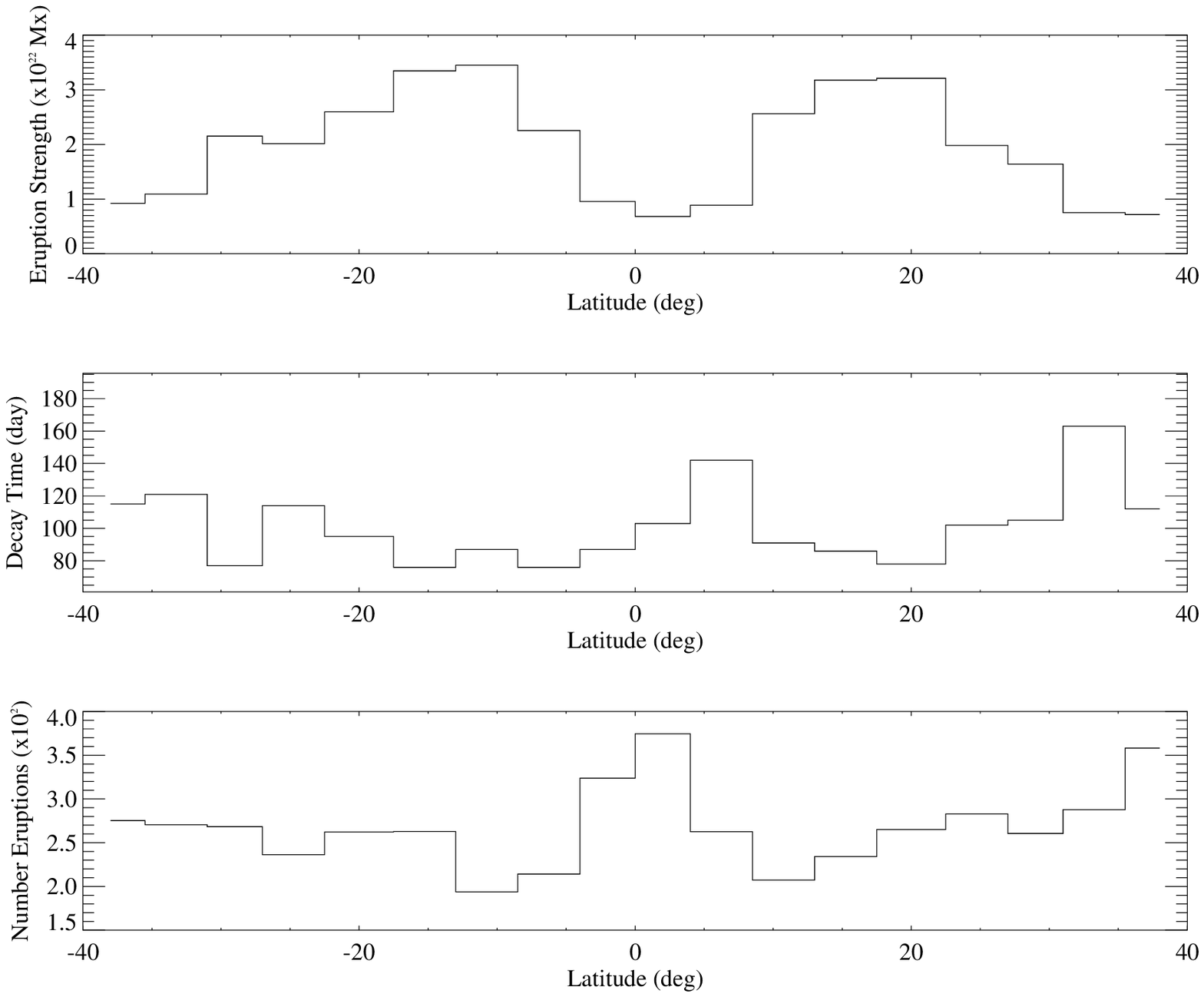}
\caption{Emergence strength (top), decay time (middle), 
and number of eruptions (bottom) per latitude bin for the magnetic activity
model used to create the simulated time series. These values were
estimated by minimizing the standard deviation between the filtered- and 
modeled-flux time series.}
\end{figure}

The modeled flux emergence is governed by the input filtered-flux time series. After
an initial period of approximately one hundred days of flux emergence to insure 
that equilibrium has been reached with the filtered-flux, the modeled time series
is created for the same interval and cadence as the observed data set used
to create {Figure~1}. Each latitude bin is assumed to rotate in 
agreement with the rotational profile observed by \citet{how70}. New flux 
is injected into a given latitude bin if the model flux drops below the 
filtered-flux time series for that latitude.
The eruptions occur randomly in longitude over a range between 0 and 2$\pi$. The 
eruption strengths are random with a distribution and mean value estimated from 
the filtered-flux time series. After a period of five times the decay-time the
contribution from any given flux emergence is set to zero. 
For the 8839-day interval simulated, there were 
approximately $4.84 \times 10^{4}$ eruptions distributed over the 18 latitude 
bins for each model realization.

\begin{figure}
\plotone{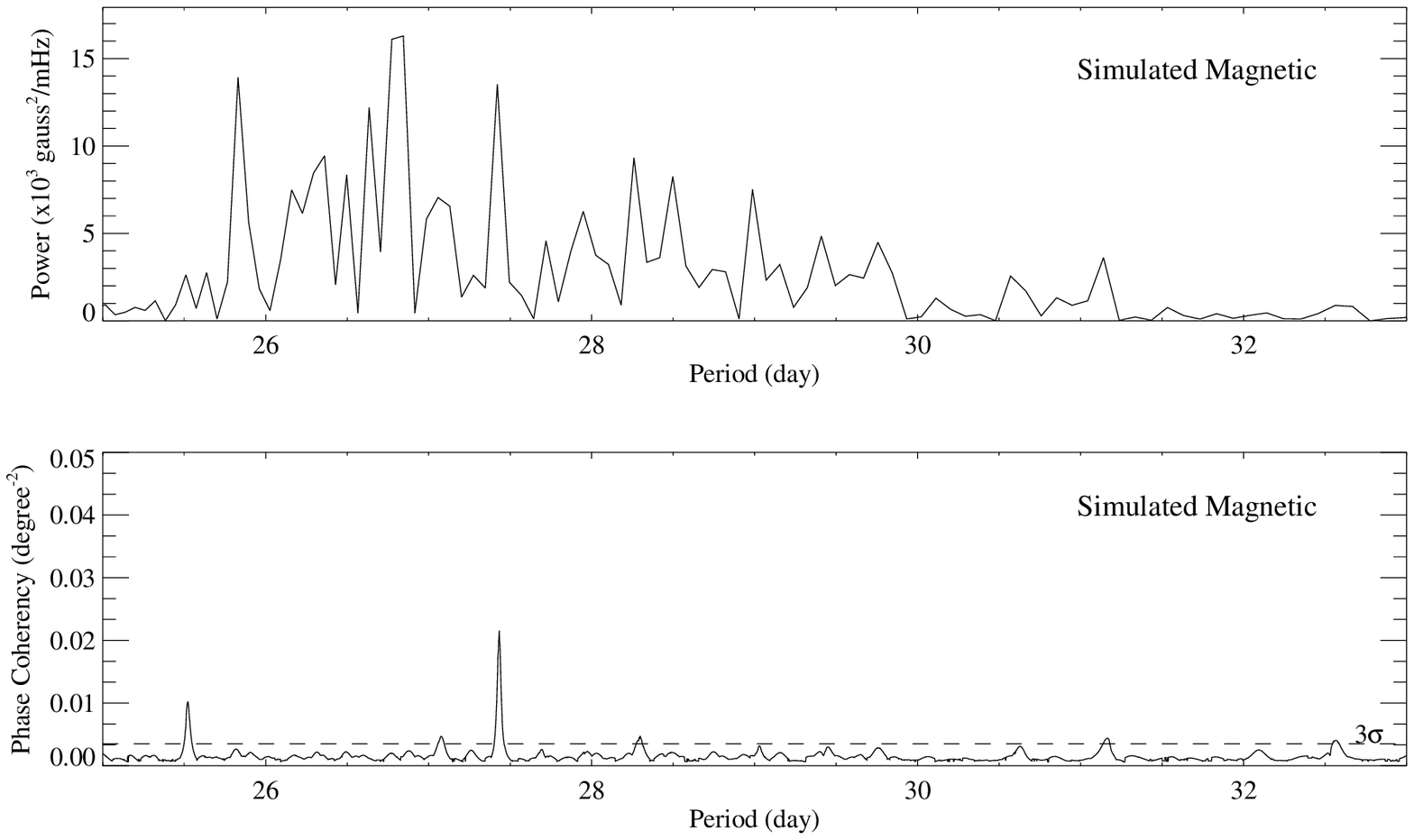}
\plotone{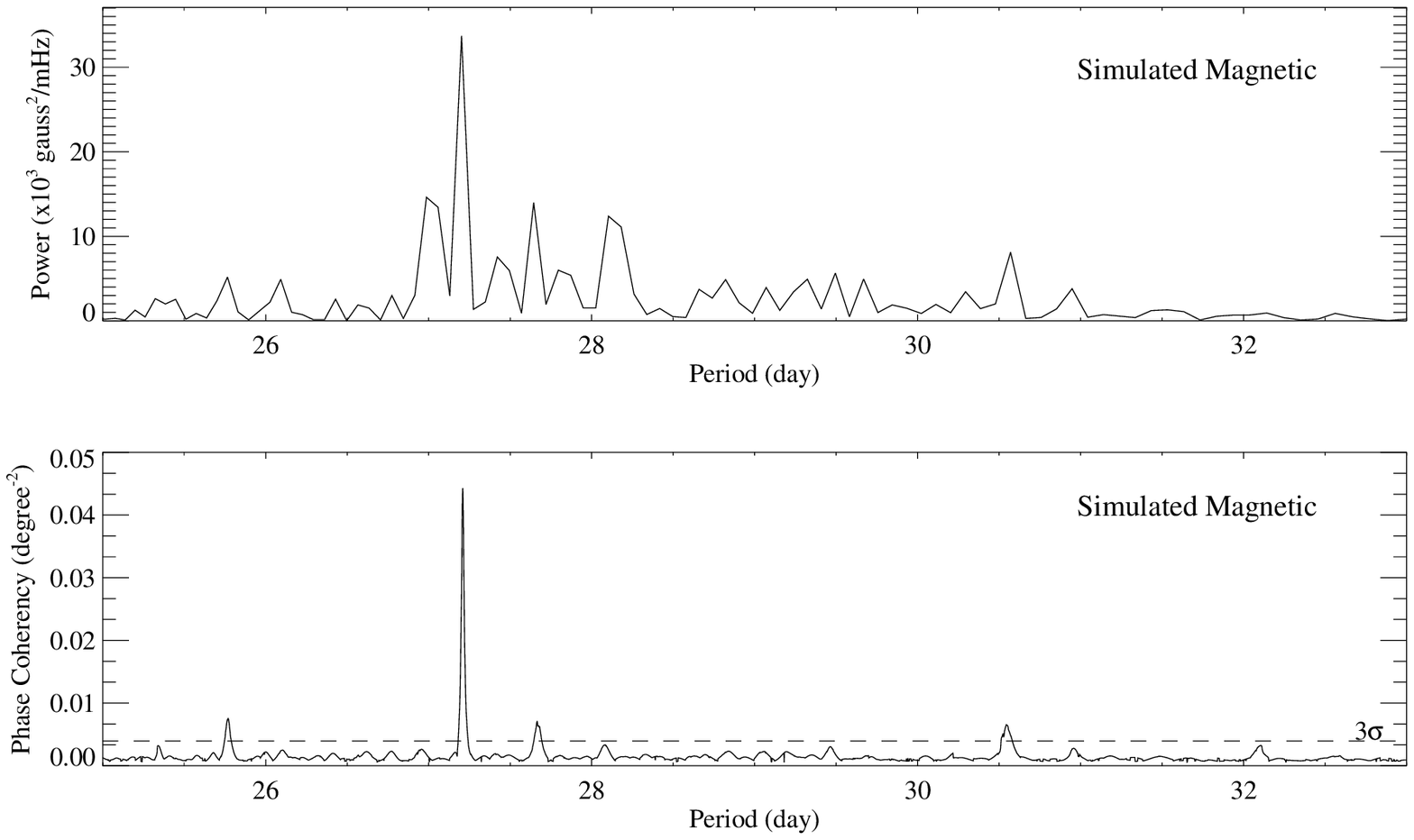}
\caption{Comparison of power (top and second from the bottom) and 
phase coherence (bottom and second from the top) spectra between 
two sample simulated unsigned magnetic flux integrated full-disk 
time series. Time series with phase coherence values of approximately 
0.02 (top two) and 0.04 (bottom two) respectively are exhibited.
The 3-$\sigma$ noise level for the phase coherence spectra
is delineated by the horizontal dashed lines.
Each simulated time series is for the same interval and cadence
as in Figure 1.}
\end{figure}

\begin{figure}[!t]
\plotone{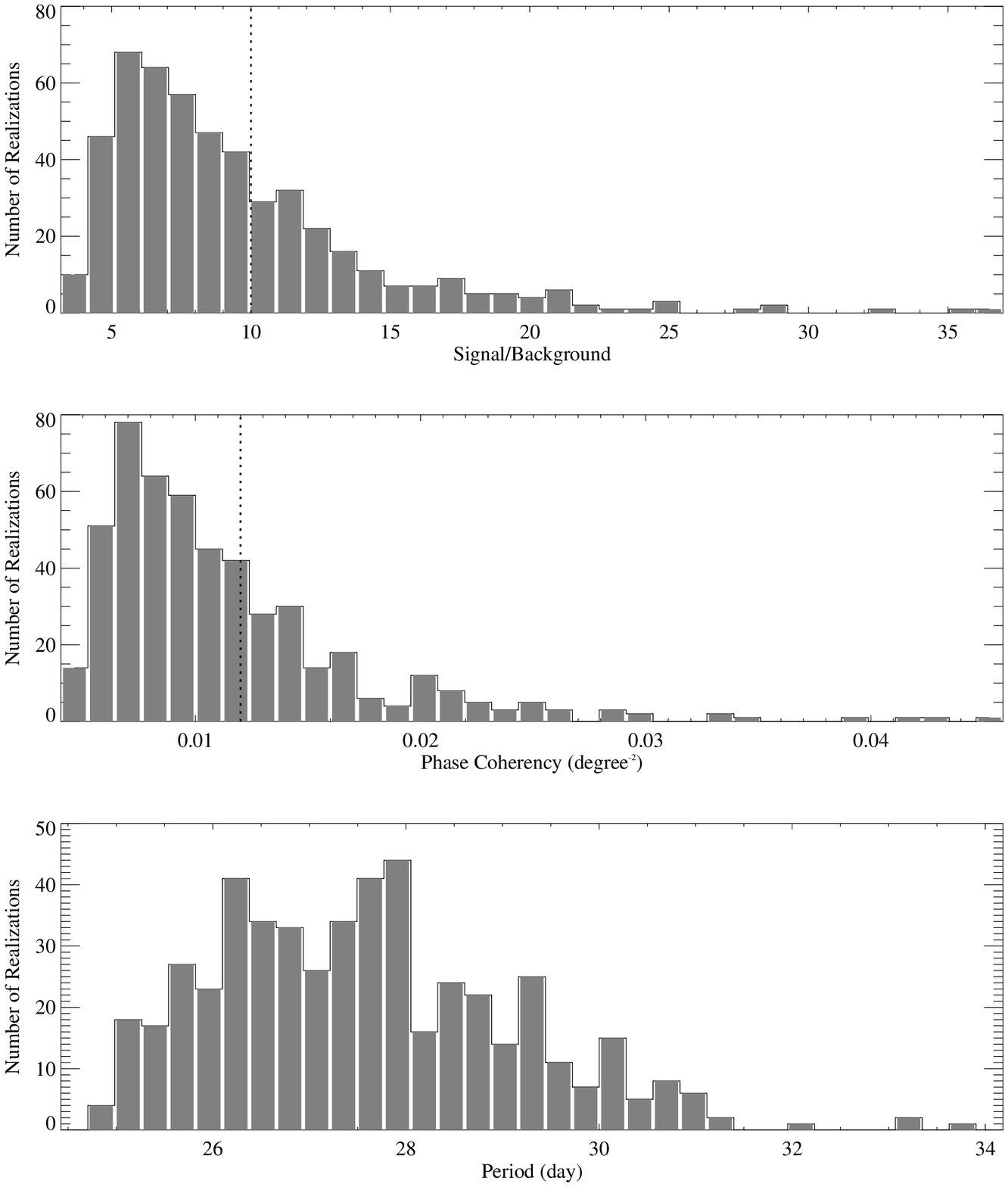}
\caption{Each of the 500 realizations of the modeled solar magnetic 
activity time series, the signal having the maximum phase coherence is 
selected. For these signals the distribution of signal-to-background (top),
phase coherency (middle), and  period (bottom) of the maximum phase 
coherence spectrum signal are shown. The vertical dotted line illustrates 
a signal-to-background ratio of 10 (top) and a phase coherence 
of 0.012 (middle). See text for discussion.}
\end{figure}

\section{Modeled Time Series Analysis}
The solar magnetic activity model was run for 500 realizations, each case 
creating a simulated integrated full-disk magnetic flux time series with a
total of 8839 days at a cadence of 1 day. 
Each time series was tested for signals that show similar 
levels of coherence to those illustrated in {Figure~1}, where a fixed 
frequency harmonic signal is considered coherent if it has negligible temporal 
variance in phase for the interval of time investigated. Following 
\citet{hen02}, we estimate the phase and amplitude of a signal from a non-linear 
least-squares fit of a fixed frequency cosine function for
overlapping segments of a time series. The phase can also be obtained with
Fourier analysis, with the caveat that the phase error estimate is not 
readily available \citep[for more discussion, see][]{hen02}. Examples of the power and 
phase coherence spectra for the simulated unsigned magnetic flux integrated 
full-disk time series are shown in Figure~3. The top two and bottom two 
figures, from simulated time series with phase coherence values of 
approximately 0.02 and 0.04 respectively, are similar to those shown in Figure~1.

\section{Results and Discussion}
As a measure of the similarity between the simulated time series and those
observed, the signal having the maximum phase coherence is selected for each 
of the 500 realizations. The distribution of the measured 
signal-to-background (S/B), the phase coherency, and period of the maximum 
phase coherence spectrum signal from each realization of the modeled solar 
magnetic activity are shown in Figure~4. In \citet{hen02}, it 
was shown that a phase coherence value greater than 0.012 or with a S/B
greater than 10 for the length of time investigated here corresponds to a signal 
that is phase coherent for approximately over 50\% of the interval investigated. 
Illustrated by the vertical dotted lines in Figure~4, roughly 33\% of 
the maximum phase coherent signals have a S/B greater than 10. 
In addition, nearly 34\% of the signals have a phase coherence
value greater than 0.012. By comparison, the 27.03-day signal in the sunspot 
number time series in {Figure~1} has phase coherence and S/B values
of approximately 0.016 and 7 respectively. 

If the solar magnetic model also included activity ``nesting'' on time scales 
up to 7 months, as reported by \citet{detoma00}, we would expect coherence values 
approaching those of the observed magnetic time series. This is supported with an 
extreme test using the current model in a mode that allows for coherent eruptions at
a fixed period of rotation for a chosen latitude, while still matching the observed mean 
flux for each latitude bin as discussed in Section~2. This resulted in notably higher 
than observed peaks at the selected rotational period in both the power and coherence 
spectra. Observationally, the 27.03-day signal shown in Figure~1 is found to be phase 
coherent for durations of three years for terms scattered throughout the 24-year 
time series \citep[see Figure~2 in][]{hen02}. This suggests that activity nesting may be 
sustained to form long lived magnetic complexes (LLMC) lasting for periods of a few years. 
This tentatively suggests that if the LLMC are the explanation for the observed coherence, 
they are scattered throughout the entire time series. If this is the case, the LLMC must rotate 
with the same period to contribute to the coherence spectrum. Tentative arguments favoring 
this suggestion can in fact be advanced: LLMC could resist dispersion if they have not broken 
their connection to the lower solar convection zone. Conceivably then, the generation of such 
fields can be favored at a given latitude. Work aimed at clarifying these issues is 
in progress.

\section{Conclusion}
We have shown that random eruptions of magnetic flux, 
modulated with solar differential rotation, have a surprisingly high 
probability of resulting in a coherent signal comparable to observation.
The solar magnetic model utilized here with longitudinally random eruptions 
is treated as a baseline case since additional processes must be included 
to explain observed nesting of magnetic activity on time scales of months to a
few years. For example, in \citet{hen02} it was found that the origin of the 27.03-day 
signal is most likely the result of long-lived, on the order of a few years,
complexes of magnetic activity in the solar northern hemisphere.
Whether long-lived complexes of magnetic activity merely by themselves are 
sufficient to explain the observed periodicities is under study. The inclusion 
of activity complexes or nesting, e.g. \cite{Schrij03}, with variable lifetimes 
and latitudes is under study and will be presented in future work. With the inclusion 
of nesting processes in the model, it is expected that the probability of resulting 
coherent signals would increase. This strongly suggests that spectral analysis 
results with integrated full-disk parameter signals, e.g. the sunspot number 
time series, should be interpreted with care. It should be stressed that 
these results do not lessen the possibility of very long-lived (between 30 to 300 
solar rotations) magnetically active longitudes. However, these findings 
reinforce the need for spatial source tracking and phase coherence analysis 
to better understand the source of these signals and substantiate claims of 
very long-lived active longitudes resulting from spatially integrated time 
analysis of solar parameters.

\section{Acknowledgments}
The authors thank J. Harvey for his suggestions and comments with regards 
to this work. National Solar Observatory (NSO) Kitt 
Peak data used here are produced cooperatively by NSF/AURA, NASA/GSFC, 
and NOAA/SEL. This research was supported in part by the Office of Naval 
Research Grant N00014-91-J-1040. The NSO is operated by the Association of 
Universities for Research in Astronomy, Inc. under cooperative agreement 
with the National Science Foundation.

\end{document}